\pdfoutput=1
\documentclass[11pt]{article}

\usepackage[]{ACL2023}

\usepackage{times}
\usepackage{latexsym}

\usepackage[T1]{fontenc}
\usepackage[utf8]{inputenc}

\usepackage{microtype}

\usepackage{inconsolata}
\usepackage{enumitem}
\usepackage{booktabs}
\usepackage{multirow}
\usepackage{colortbl}
\usepackage{xcolor}
\usepackage{array} 
\usepackage{subfigure}
\usepackage[graphicx]{realboxes}
\usepackage{tikz}
\usepackage{amsmath, amssymb}
\usepackage{adjustbox}
\usepackage{xspace}
\usepackage{hyperref}

\usepackage{bbding}
\usepackage{soul} % 导入 soul 包
\usepackage{color}
\usepackage{arydshln}
\usepackage{float}
\title{$\mathbb{USCD}$: Improving Code Generation of LLMs by Uncertainty-Aware Selective Contrastive Decoding}

\author{Shuai Wang\textsuperscript{\rm 1}\space\space
Liang Ding\textsuperscript{\rm 2}\space\space
Li Shen\textsuperscript{\rm 3}\space\space
Yong Luo\textsuperscript{\rm 1}\space\space
Zheng He\textsuperscript{\rm 1}\space\space
Wei Yu\textsuperscript{\rm 1}\space\space
Dacheng Tao\textsuperscript{\rm 4}\\
    \textsuperscript{\rm 1}Wuhan University 
    \hspace{5em}\textsuperscript{\rm 2}The University of Sydney \\
   \textsuperscript{\rm 3}Sun Yat-sen University
    \hspace{2em}\textsuperscript{\rm 4}Nanyang Technology University\\
    {wangshuai123@whu.edu.cn},
    {liangding.liam@gmail.com}
}

\begin{document}
\maketitle
\begin{abstract}
Large language models (LLMs) have shown remarkable capabilities in code generation. However, the effects of hallucinations (e.g., output noise) make it particularly challenging for LLMs to generate high-quality code in one pass. In this work, we propose a simple and effective \textbf{u}ncertainty-aware \textbf{s}elective \textbf{c}ontrastive \textbf{d}ecoding ($\mathbb{USCD}$) mechanism to improve the quality of one-pass code generation in LLMs and reduce the impact of output noise. To be specific, we first elaborately designed a negative prompt (namely lame prompt) to output noise by removing input-output examples from the standard few-shot prompt. Our preliminary study shows that the Jensen-Shannon divergence (JS divergence) between token distribution uncertainty and the output noise is relatively low (approximately $0.25$), indicating their high relevance. Then, we selectively eliminate output noise induced by lame prompts based on the uncertainty of the prediction distribution from the standard prompt. Notably, our proposed plug-and-play mechanism is an inference-only method, enjoying appealing flexibility. Extensive experiments on widely used benchmarks, e.g., HumanEval, MBPP, and MultiPL-E, upon several LLMs (i.e., Inocder-6b, CodeLlama-7b, WizardCoder-15b, StarCoder, and Llama2-7b), demonstrate that our proposed USCD significantly improves one-pass code generation, with an average \textit{pass@$1$} scores increase of 16.59\%. We will release code and data on GitHub.
\end{abstract}

\section{Introduction}
Large language models (LLMs,~\citealp[]{openai2023gpt4, touvron2023llama}) have achieved widespread success across many NLP tasks~\cite{zhong2023chat,Peng2023ChatGPT4MT,ren2024healthcare} due to their remarkable emergent abilities~\cite{DBLP:journals/tmlr/WeiTBRZBYBZMCHVLDF22}. One of the most exciting emergent abilities is code generation~\cite{rozière2023code,khojah2024beyond}, which aims at producing the executable code based on user prompts (i.e., standard prompts). 
\begin{figure}[!t]
    \centering
    \includegraphics[width=0.47\textwidth]{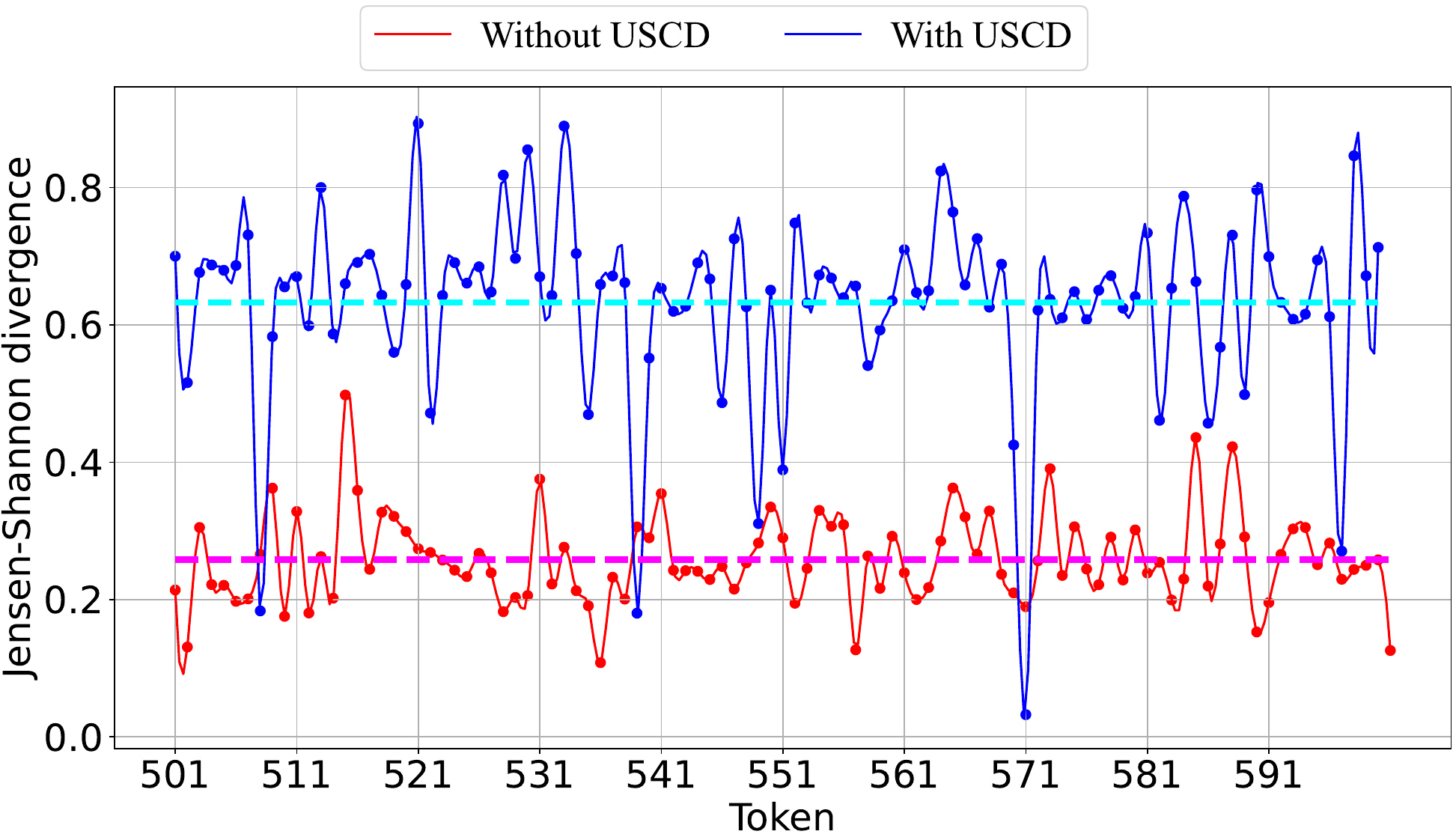}
    \caption{\textbf{The Jensen-Shannon divergence (JS divergence) between token distribution uncertainty and output noise for Incoder-6b~\cite{fried2023incoder}}. We randomly selected a standard prompt that generated incorrect code with Incoder-6b, i.e., $501 \sim 600$ tokens in HumanEval/163. We calculated the JS divergence between the token distribution of the lame prompt output and the token distribution with (blue) and without (red) the USCD mechanism. We can clearly see that for the Incoder-6b, the JS divergence between token distribution uncertainty and output noise is approximately $0.25$ without using the USCD mechanism (red) and approximately $0.65$ with the USCD mechanism (blue).}
    \label{fig:JS_divergence}
\end{figure}
While LLMs have shown excellent abilities in natural language tasks,
% and $32.60$ on BBH~\cite{suzgun2022challenging},
% GPT-4.0~\cite{openai2023gpt4} achieved scores of $86.40$ on MMLU and $95.30$ on HellaSwag~\cite{zellers2019hellaswag},
their performance of code generation in one pass through standard prompts is often concerning, e.g., Llama2-7b~\cite{touvron2023llama} scores $45.30$ on MMLU~\cite{hendrycks2021measuring}, but only $12.80$ on HumanEval~\cite{chen2021evaluating} and even only $4.62$ on OOP~\cite{wang2024oop}.
% e.g., Llama2-7b~\cite{touvron2023llama} scored $45.30$ on MMLU~\cite{hendrycks2021measuring}, $12.80$ on HumanEval~\cite{chen2021evaluating}. 
Unlike natural languages, programming languages have strict syntax and semantic rules~\cite{naur1975programming,mandrioli2015programming}, which can easily cause LLMs to produce hallucinations (e.g., output noise) during one-pass code generation, making it particularly difficult to generate high-quality code.

% Unlike natural languages, programming languages have strict syntax and semantic rules~\cite{naur1975programming,mandrioli2015programming}, which can easily cause LLMs to produce hallucinations (e.g., output noise) during one-pass code generation, making it particularly difficult to generate high-quality code.
% Unlike natural language, programming languages have strict syntax and semantic rules~\cite{naur1975programming,mandrioli2015programming} that must be precisely followed for the computer to correctly understand and execute the code. 

\textbf{[Limitations of existing methods]}
To improve the quality of one-pass generated code~\cite{logothetis1981compiling}, most existing methods primarily focus on pre-trained or fine-tuned models~\cite{rozière2023code, li2023starcoder,luo2023wizardcoder}, and post-processing repair~\cite{yasunaga2021break, chen2023teaching, olausson2023selfrepair}. 
Although pretraining or fine-tuning models can reduce the output noise of LLMs when generating code in one pass by updating the model's parameters, it requires a large amount of corpus and computational resources. Post-processing repair methods typically use feedback information obtained from the feedback model to perform secondary or multiple rounds of repair on the initially generated results. However, post-processing repair methods do not reduce the output noise of LLMs when generating code in one pass.
Moreover, recent research~\cite{huang2023large, valmeekam2023can} indicates that post-processing repair methods cannot achieve improved results without additional external feedback information.

\begin{figure}[!t]
    \centering
    \includegraphics[width=0.47\textwidth]{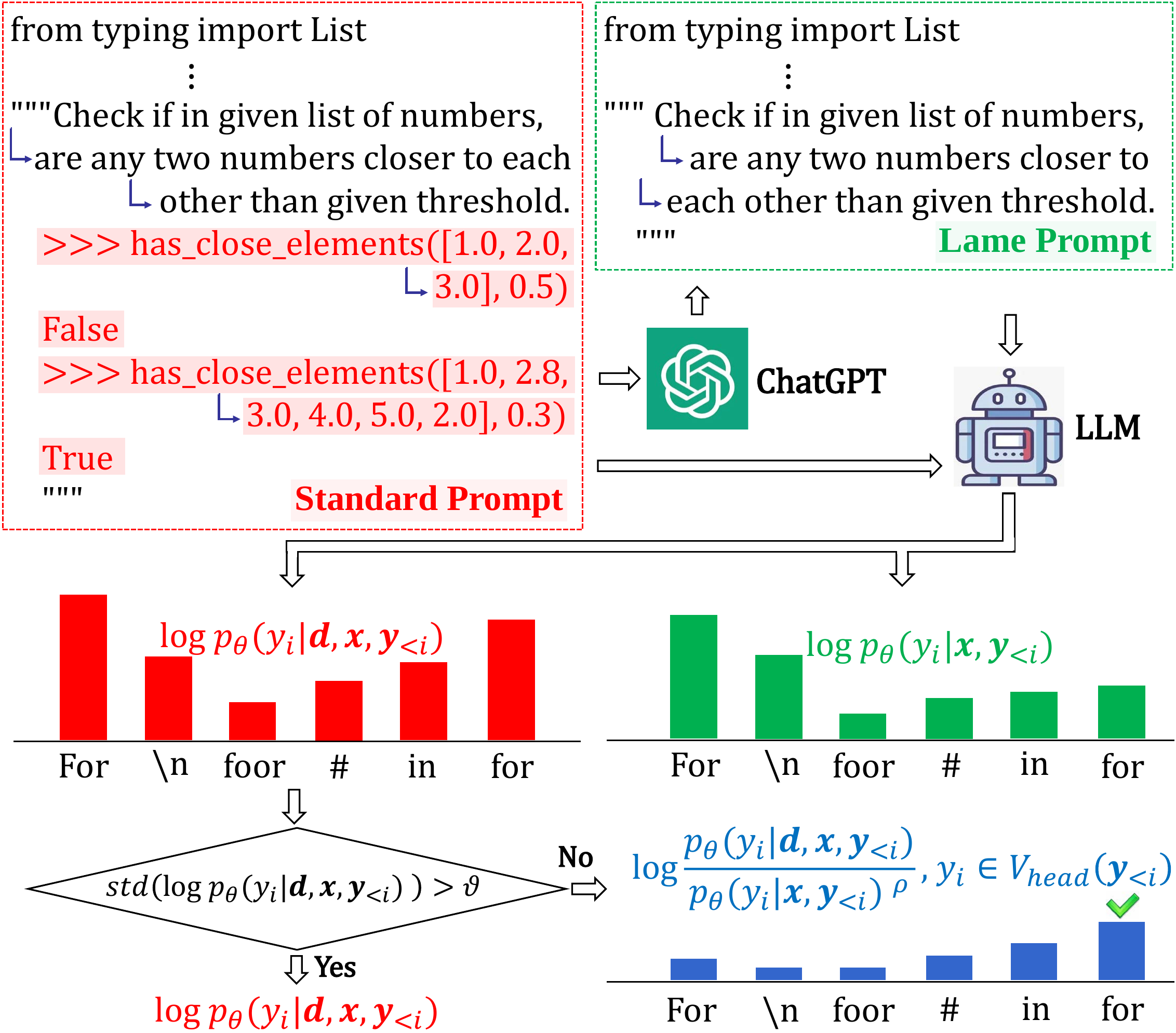}
    \caption{\textbf{Illustration of our uncertainty-aware selective contrast decoding (USCD) mechanism for improving code generation of LLMs}.}
    \label{fig:motivate}
\end{figure}

\textbf{[Motivation]} 
Therefore, we are considering whether leveraging information from standard prompts can lead to more accurate one-pass code generation and mitigate output noise, all without requiring model parameter updates. To this end, we carefully designed a lame prompt to generate output noise by removing input-output examples from the standard prompt. Our preliminary study indicates that the JS divergence between token distribution uncertainty and output noise is close to $0.25$ (as illustrated in Figure~\ref{fig:JS_divergence}), illustrating a high correlation.

\begin{figure}[!t]
    \centering
    \includegraphics[width=0.47\textwidth]{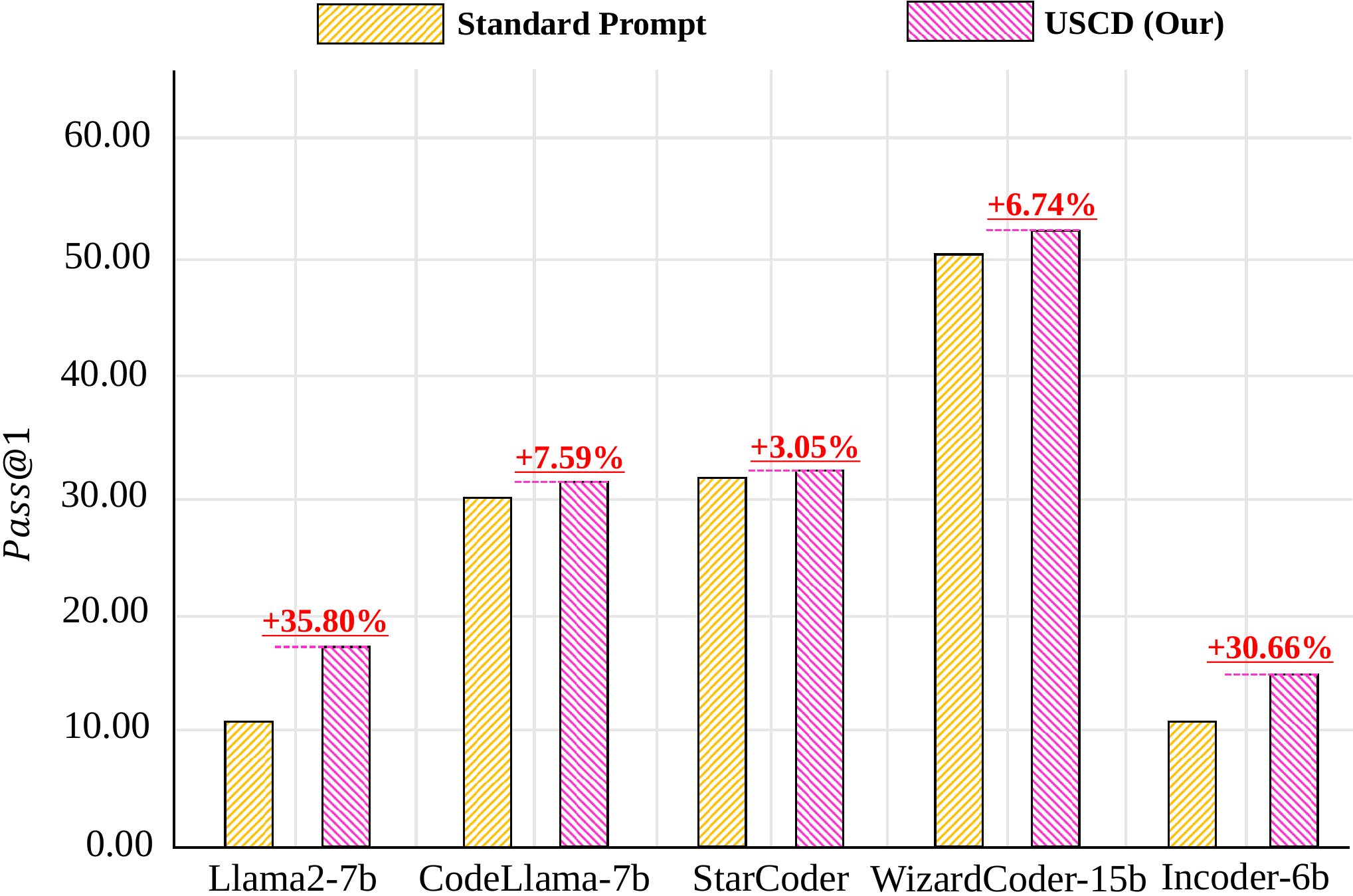}
    \caption{\textbf{Comparison of the performance between models using the USCD mechanism and models directly using standard prompts on the HumanEval benchmark~\cite{chen2021evaluating}}. During the experiment, we use a temperature of $0.1$ and top-$p$=$0.95$. We can see that USCD mechanism significantly improves the performance of code specialized models, e.g., CodeLlama-7b~\cite{rozière2023code}, StarCoder~\cite{li2023starcoder}, WizardCoder~\cite{luo2023wizardcoder}, Incoder-6b~\cite{fried2023incoder} and general models, e.g., Llama2-7b~\cite{touvron2023llama} alike.}
    \label{fig:result_illustrated}
\end{figure}

\textbf{[Method]} Motivated by this, we propose a novel uncertainly-aware selective contrastive decoding (USCD) mechanism, as illustrated in Figure~\ref{fig:motivate}. Our USCD mechanism operates by initially using the standard deviation to prejudge the presence of noise in the logit of the standard prompt. Then, for the current standard prompt identified with noise, it applies the logit of the lame prompt to correct it, thereby achieving the goal of enhancement code generation result. Encouragingly, our preliminary experiments in Figure~\ref{fig:JS_divergence} and~\ref{fig:result_illustrated} show that our USCD mechanism effectively reduces the impact of output noise and significantly improves the performance of one-pass code generation.

\textbf{[Contributions]} Our main contributions are:

\begin{itemize}
\item We meticulously devise a lame prompt to induce the noise present in a standard prompt generation. The construction of the lame prompt does not require intervention from external knowledge.

\item To elicit the induced noises, we then design an uncertainly-aware selective contrastive decoding (USCD) mechanism to improve the code generation for LLMs. 
% To the best of our knowledge, we are the first to utilize contrastive decoding to improve one-pass code generation.

% \item The proposed USCD is flexible and scalable and can be applied to any LLMs for better one-pass code generation. Moreover, the USCD mechanism does not involve updating the model parameters and secondary or multiple operations.

\item Extensive experiments have shown that our flexible and scalable USCD significantly and consistently improves the precision of LLMs in generating code in one-pass, with an average score increase of up to 16.59\% in \textit{pass@$k$}.
\end{itemize}

\section{Methodology}
\label{sec:method}
\subsection{Overview}
Given a LLM $\theta$, a natural language description $\boldsymbol{x}$, and input-output examples $\boldsymbol{d}$, the process of generating the corresponding code using LLM is:
\begin{equation}
\label{eq:generating_code_process}
y_i \sim p_\theta \left(y_i \mid \boldsymbol{d}, \boldsymbol{x}, \boldsymbol{y}_{<i}\right),
\end{equation}
where $y_i$ denotes the token at time step $i$, and $\boldsymbol{y}_{<i}$ represents the sequence of generated tokens up to the time step $(i-1)$.

However, the LLM $\theta$ does not always accurately predict the maximum logit value (i.e., $max(p_\theta (y_i \mid \boldsymbol{d}, \boldsymbol{x}, \boldsymbol{y}_{<i}))$) for token at time step $i$. This can lead to errors in the code for one-pass generation, e.g., when the LLM $\theta$ generates the corresponding ``\textit{for}'' code based on ``\textit{Check if in the given list of numbers, are any two numbers closer to each other than given threshold}'', and the input-output examples ``\textit{$>>>$has\_close\_elements([1.0, 2.0, 3.0], 0.5)$\backslash$n $\quad$ False$\backslash$n   $>>>$ has\_close\_elements([1.0, 2.8, 3.0, 4.0, 5.0, 2.0], 0.3)$\backslash$n $\quad$ True$\backslash$n}'', it erroneously predicts ``\textit{For}''. We refer to the probability distribution that generates incorrect $max(p_\theta (y_i \mid \boldsymbol{d}, \boldsymbol{x}, \boldsymbol{y}_{<i}))$ as the probability distribution of code noise. Although ``for'' is capitalized as ``For'', it does not run normally when tested using the evaluator, as demonstrated in Figure~\ref{fig:motivate}.

If the current noise, i.e., $max(p_\theta (y_i \mid \boldsymbol{d},\boldsymbol{x},\boldsymbol{y}_{<i}))$ value, is eliminated during the process of generating logits according to standard prompts, it can improve the accuracy of generating code at once, as demonstrated in Figure~\ref{fig:motivate}. 
Therefore, we carefully constructed a lame prompt by removing input-output examples $\boldsymbol{d}$ from the standard prompt, generating a stable and completely noisy logit distribution. The construction process of the lame prompt is detailed in section~\ref{sec:noise_prompt_construction}. However, the maximum logits value generated by LLMs doesn't always necessarily entail noise (i.e., the error of $max(p_\theta (y_i \mid \boldsymbol{d}, \boldsymbol{x}, \boldsymbol{y}_{<i}))$). To this end, we propose a novel uncertainty-aware selective contrastive decoding (USCD) mechanism to improve the accuracy of one-pass generating code in LLMs. 

\subsection{Construction of the Lame Prompt}
\label{sec:noise_prompt_construction}
\begin{figure}[!t]
    \centering
    \includegraphics[width=0.47\textwidth]{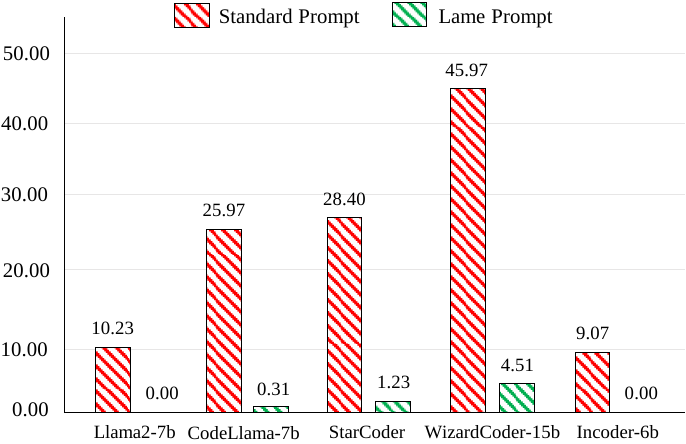}
    \caption{\textbf{Performance comparison of the used LLMs, e.g., Llama2-7b~\cite{touvron2023llama}, CodeLlama-7b~\cite{rozière2023code}, and StarCode~\cite{li2023starcoder}, using standard prompt and lame prompt on the HumanEval benchmark~\cite{chen2021evaluating}}. We can clearly see that the performance of LLMs using a lame prompt is significantly lower compared to using a standard prompt.}
    \label{fig:compcare_general_noise}
\end{figure}

The lame prompt (aka. negative prompt in the USCD mechanism) is a crucial component of the USCD mechanism and forms a probability distribution with inherent noise at the inference stage of the LLM $\theta$. According to Eq.~(\ref{eq:generating_code_process}), the LLM $\theta$ strongly relies on input-output examples $\boldsymbol{d}$ during the inference process. If the LLM $\theta$ does not excessively focus on input-output examples $\boldsymbol{d}$ and instead relies on external knowledge, it will struggle to generate correct code in one-pass, as illustrated in Figure~\ref{fig:compcare_general_noise}. Our constructed lame prompts, when inference through LLMs, can generate stable and fully noisy logit distributions.

In this work, we construct a standard prompt and its corresponding lame prompt as a few-shot example, enabling an LLM to reference a few-shot example to remove input-output examples $\boldsymbol{d}$ of standard prompts from HumanEval~\cite{chen2021evaluating}, MBPP~\cite{austin2021program}, and MultiPL-E~\cite{multiple} benchmarks~\footnote{Note: The lame prompt method we adopted is just one of many approaches.}. The specific construction process of the lame prompt is shown in Appendix~\ref{sec:lame_prompt}.

\subsection{Uncertainly-Aware Selective Contrastive Decoding}
\label{sec:uscd_mechanism}
\textbf{Prejudgment of standard deviation.}
Utilizing the standard deviation $\mu$ to measure the dispersion of the probability distribution $\boldsymbol{y}_{<i}$ (which can be seen as one type of estimation of uncertainty\footnote{While various uncertainty estimation methods exist, such as computing semantic entropy~\cite{farquhar2024detecting} and using larger models~\cite{wang-etal-2024-uncertainty}, our standard deviation approach offers a simple yet effective alternative, akin to simplified version of semantic entropy, estimated directly from the model itself rather than costly annotation by larger models, e.g., GPT-4.} we can pre-judge the degree of noise in the current probability distribution, i.e., whether the LLM $\theta$ has generated correct code:
\begin{equation}
\label{eq:standard_deviatio_judgment}
\begin{split}
&\max(p_\theta (y_i \mid \boldsymbol{d}, \boldsymbol{x}, \boldsymbol{y}_{<i}))=\begin{cases}
correct,
& \mu > \vartheta \\
error,    	      & \text{otherwise}
\end{cases}, \\
&where \, \mu=\sqrt{\frac{1}{n} \left(y_i\in V:p_\theta (y_i \mid \boldsymbol{d}, \boldsymbol{x}, \boldsymbol{y}_{<i})-\overline{y_i}\right)^2} 
\end{split}
\end{equation}
\noindent
Where, $V$ represents the output vocabulary in the LLM $\theta$, $\vartheta$ denotes the threshold, $n$ represents the length of the output vocabulary, and $\overline{y_i}$ denotes the average prediction of token probability, i.e., $\overline{y_i}=mean\left(p_\theta \left(y_i \mid \boldsymbol{d}, \boldsymbol{x}, \boldsymbol{y}_{<i}\right)\right)$, at time step $t$.

\noindent
\textbf{Rationality constraint.}
For the probability distribution $p_\theta \left(y_i \mid \boldsymbol{d}, \boldsymbol{x}, \boldsymbol{y}_{<i}\right)$ with the standard prompt containing noise, it is necessary to employ the lame prompt induction constructed in Section~\ref{sec:noise_prompt_construction}. We follow the rationale constraint filtering approach proposed by~\citet{li2023contrastive}, filtering out smaller logit values from the probability distribution $p_\theta \left(y_i \mid \boldsymbol{d}, \boldsymbol{x}, \boldsymbol{y}_{<i}\right)$, i.e.,
\begin{equation}
\label{eq:rationality_constraint}
V_{thresh} =\{y_i\in V:p_\theta \left(y_i \mid \boldsymbol{d}, \boldsymbol{x}, \boldsymbol{y}_{<i}\right) \geq \eta \cdot \overline{y_i}\}
\end{equation}

\noindent
Here, $\eta$ is a hyperparameter, which we set to $0.1$ following~\citet{li2023contrastive}.

Due to significant differences between programming languages and natural languages, and considering that the mean value $\overline{y_i}$ is a statistical measure describing the central location in a probability distribution. 
Unlike the CD~\cite{li2023contrastive}, we use the mean value $\overline{y_i}$ to filter out the smaller logit values in the probability distribution $p_\theta \left(y_i \mid \boldsymbol{d}, \boldsymbol{x}, \boldsymbol{y}_{<i}\right)$.

\noindent
\textbf{The code generation mechanism based on uncertainty-aware selective contrastive decoding}. Combining the constraints of rationality, we use uncertainty-aware selective contrastive decoding to eliminate noise in the probability distribution of standard prompt reasoning, i.e.,
\begin{equation}
\label{eq:contrastive_decoding_repair}
% \begin{split}
score_{cd}\left(i\right) = 
\begin{cases}
\log \frac{\left(y_i \mid \boldsymbol{d}, \boldsymbol{x}, \boldsymbol{y}_{<i}\right)}{\left(y_i \mid \boldsymbol{x}, \boldsymbol{y}_{<i}\right)^{\rho}},
& if \, V_{thresh}\left(\boldsymbol{y}_{<i}\right) \\
-inf, & \text{otherwise} 
\end{cases} 
% \end{split}
\end{equation}

By employing the uncertainty-aware selective contrastive decoding $score_{cd}\left(i\right)$, we eliminate noise in the probability distribution  of standard prompts, addressing errors in code syntax, semantics, and other aspects that may occur during the one-pass code generation process.

\section{Experiments}
\label{sec:experiment}
\subsection{Experimental Setup}
\noindent
\textbf{Datasets.}
We follow the research of~\cite{rozière2023code, touvron2023llama, li2023starcoder, du2022glm} and have selected three benchmarks, e.g., HumanEval~\cite{chen2021evaluating}, MBPP~\cite{austin2021program} and MultiPL-E~\cite{multiple}, to validate the performance of USCD mechanism. The detailed description of HumanEval, MBPP and MultiPL-E benchmarks is shown in Appendix~\ref{sec:test_benchmark}.

\noindent
\textbf{Models.}
To better demonstrate the performance of the USCD mechanism, we select general models, e.g., Llama2-7b~\cite{touvron2023llama} and code-specialized models, e.g., CodeLlama-7b~\cite{rozière2023code}, StarCode~\cite{li2023starcoder}, WizardCoder-15b~\cite{luo2023wizardcoder}, Incoder-6b~\cite{fried2023incoder}. The details of these models are shown in Appendix~\ref{sec:model}.

\noindent
\textbf{Evalution metrics.}
We follow~\cite{rozière2023code, touvron2023llama, li2023starcoder,luo2023wizardcoder} and use the \textit{pass@$k$} metric~\cite{chen2021evaluating} to evaluate the improved capability of the USCD mechanisms. The \textit{pass@$k$} metric is calculated by testing the pass rate of the currently generated code using test cases, i.e.,
\begin{equation}
\label{eq:mertric_pass@k}
\textit{pass@$k$}:=\mathop{\mathbb{E}}_{Problems} \left[1-\frac{{\binom{n-c}{k}}}{\binom{n}{k}} \right]
\end{equation}
In Eq. (\ref{eq:mertric_pass@k}), $n$ represents the number of code generations for a given problem; $c$ represents the quantity of $n$ generated codes passing tests. In the experiment, We evaluate the USCD mechanism on eight NVIDIA A100 GPUs using the bigcode framework~\footnote{\url{https://github.com/bigcode-project/bigcode-evaluation-harness}}. Besides, we set $k=15$.

\begin{table*}[!t]
  \centering
  \resizebox{1.0\linewidth}{!}{
    \begin{tabular}{ccccccccc}
    \toprule
    \multirow{2}[4]{*}{\textbf{Reduced number of $\boldsymbol{d}$}} & \multicolumn{7}{c}{\textbf{CodeLlama-7b}}                      & \multirow{2}[4]{*}{Avg} \\
    \cmidrule{2-8}          & \textit{pass@$1$} & \textit{pass@$3$} & \textit{pass@$5$} & \textit{pass@$8$} & \textit{pass@$10$} & \textit{pass@$12$} & \textit{pass@$15$} &  \\ 
    \midrule
    $1$     & $1.02_{\textcolor{blue}{\textbf{(-24.95)}}}$  & $2.21_{\textcolor{blue}{\textbf{(-39.47)}}}$  & $2.95_{\textcolor{blue}{\textbf{(-46.64)}}}$  & $3.73_{\textcolor{blue}{\textbf{(-53.27)}}}$  & $4.14_{\textcolor{blue}{\textbf{(-56.28)}}}$  & $4.48_{\textcolor{blue}{\textbf{(-58.56)}}}$  & $4.88_{\textcolor{blue}{\textbf{(-60.97)}}}$  & $3.34$  \\
    $2$     & $0.41_{\textcolor{blue}{\textbf{(-25.56)}}}$  & $1.20_{\textcolor{blue}{\textbf{(-40.48)}}}$  & $1.97_{\textcolor{blue}{\textbf{(-47.62)}}}$  & $3.09_{\textcolor{blue}{\textbf{(-53.91)}}}$  & $3.80_{\textcolor{blue}{\textbf{(-56.62)}}}$  & $4.49_{\textcolor{blue}{\textbf{(-58.55)}}}$  & $5.49_{\textcolor{blue}{\textbf{(-60.36)}}}$  & $2.92$  \\
    $3$     & $0.61_{\textcolor{blue}{\textbf{(-25.36)}}}$  & $1.49_{\textcolor{blue}{\textbf{(-40.19)}}}$  & $2.17_{\textcolor{blue}{\textbf{(-47.42)}}}$  & $3.05_{\textcolor{blue}{\textbf{(-53.95)}}}$  & $3.60_{\textcolor{blue}{\textbf{(-56.82)}}}$  & $4.13_{\textcolor{blue}{\textbf{(-58.91)}}}$  & $4.88_{\textcolor{blue}{\textbf{(-60.97)}}}$  & $2.85$  \\
    $4$     & $0.37_{\textcolor{blue}{\textbf{(-25.60)}}}$  & $0.86_{\textcolor{blue}{\textbf{(-40.82)}}}$  & $1.19_{\textcolor{blue}{\textbf{(-48.40)}}}$  & $1.58_{\textcolor{blue}{\textbf{(-55.42)}}}$  & $1.83_{\textcolor{blue}{\textbf{(-58.59)}}}$  & $2.07_{\textcolor{blue}{\textbf{(-60.97)}}}$  & $2.44_{\textcolor{blue}{\textbf{(-63.41)}}}$  & $1.48$  \\
    \bottomrule
    \end{tabular}%
   }
  \caption{\textbf{The performance of CodeLlama-7b under different numbers of input-output examples $\boldsymbol{d}$ in the HumanEval benchmark}. The blue colour shows the difference in performance between prompts with reduced input-output examples $\boldsymbol{d}$ and the standard prompt. During the experiment, we use a temperature of $0.8$ and top-$p$=$0.95$.}
  \label{tab:different_input-output_example}%
\end{table*}%

\begin{table*}[!t]
  \centering
  \resizebox{1.0\linewidth}{!}{
    \begin{tabular}{ccccccccc}
    \toprule
    \multirow{2}[4]{*}{\textbf{Reduced number of $\boldsymbol{d}$}} & \multicolumn{7}{c}{\textbf{CodeLlama-7b}}                      & \multirow{2}[4]{*}{Avg} \\
    \cmidrule{2-8}          & \textit{pass@$1$} & \textit{pass@$3$} & \textit{pass@$5$} & \textit{pass@$8$} & \textit{pass@$10$} & \textit{pass@$12$} & \textit{pass@$15$} &  \\ 
    \midrule
    $1$  & $24.55_{\textcolor{red}{\textbf{(-2.20)}}}$  & $39.18_{\textcolor{red}{\textbf{(-3.10)}}}$  & $46.71_{\textcolor{red}{\textbf{(-3.25)}}}$  & $53.88_{\textcolor{red}{\textbf{(-3.64)}}}$  & $57.31_{\textcolor{red}{\textbf{(-4.04)}}}$  & $60.09_{\textcolor{red}{\textbf{(-4.43)}}}$  & $63.41_{\textcolor{red}{\textbf{(-4.88)}}}$  & $49.30$  \\   
    $2$ & $25.98_{\textcolor{red}{\textbf{(-0.77)}}}$  & $41.34_{\textcolor{red}{\textbf{(-0.94)}}}$  & $49.42_{\textcolor{red}{\textbf{(-0.54)}}}$  & $57.29_{\textcolor{red}{\textbf{(-0.23)}}}$  & $61.15_{\textcolor{red}{\textbf{(-0.20)}}}$  & $64.35_{\textcolor{red}{\textbf{(-0.17)}}}$  & $68.29_{\textcolor{red}{\textbf{(-0.00)}}}$  & $52.55$  \\  
    $3$ & $25.85_{\textcolor{red}{\textbf{(-0.90)}}}$  & $42.39_{\textcolor{red}{\textbf{(+0.11)}}}$  & $50.75_{\textcolor{red}{\textbf{(+0.79)}}}$  & $58.25_{\textcolor{red}{\textbf{(+0.73)}}}$  & $61.51_{\textcolor{red}{\textbf{(-0.20)}}}$  & $63.92_{\textcolor{red}{\textbf{(-0.60)}}}$  & $66.46_{\textcolor{red}{\textbf{(-1.83)}}}$  & $52.73$  \\
    $4$ & $26.09_{\textcolor{red}{\textbf{(-0.66)}}}$  & $42.75_{\textcolor{red}{\textbf{(+0.47)}}}$  & $50.91_{\textcolor{red}{\textbf{(+0.95)}}}$  & $58.30_{\textcolor{red}{\textbf{(+0.78)}}}$  & $61.69_{\textcolor{red}{\textbf{(+0.34)}}}$  & $64.07_{\textcolor{red}{\textbf{(-0.45)}}}$  & $66.75_{\textcolor{red}{\textbf{(-1.54)}}}$  & $52.94$  \\
    \bottomrule
    \end{tabular}%
   }
  \caption{\textbf{The performance of the USCD mechanism using standard prompts with gradually fewer input-output examples $\boldsymbol{d}$ as lame prompts}. The red colour shows the performance of contrastive decoding code fixing when gradually reducing the input-output examples $\boldsymbol{d}$ from standard prompts as lame prompts, compared to using standard prompts with no input-output examples $\boldsymbol{d}$ as lame prompts. During the experiment, we use a temperature of $0.8$ and top-$p$=$0.95$.}
  \label{tab:different_input-output_example_contrastive_decoding}%
\end{table*}%

\begin{table*}[!t]
  \centering
  \resizebox{0.95\linewidth}{!}{
    \begin{tabular}{cccccccccccc}
    \toprule
    \multicolumn{12}{c}{\textbf{HumanEval}} \\
    \midrule
    $\rho$     & $0.00$  & $0.10$  & $0.20$  & $0.30$  & $0.40$  & $0.50$  & $0.60$  & $0.70$  & $0.80$  & $0.90$ & $1.00$  \\
    \midrule
    Incoder-7b & $9.39 $ & $9.51$  & $9.64$  & $10.00$  & $10.08$  & $10.12$  & $10.16$  & $\textcolor{red}{\textbf{10.24}}$  & $10.08$  & 10.12 & 9.65 \\
    CodeLlama-7b & $24.51$  & $24.88$  & $25.69$  & $\textcolor{red}{\textbf{26.38}}$  & $26.10$  & $26.34$  & $25.49$  & $25.57$  & $25.37$  & $24.27$ & $23.74$ \\
    StarCoder & $28.13$  & $28.93$  & $28.90$  & $29.54$  & $29.27$  & $29.27$  & $29.38$  & $\textcolor{red}{\textbf{30.08}}$  & $29.23$  & $29.10$ & $28.35$  \\
    \bottomrule
    \end{tabular}%
  }
  \caption{\textbf{\textit{Pass@$1$} of Incoder-7b, CodeLlama-7b, and StarCoder under different values of $\rho$}. The red colour shows the best result.}
  \label{tab:coefficient_p}%
\end{table*}%

\begin{figure}[!t]
    \centering
    \includegraphics[scale=0.28]{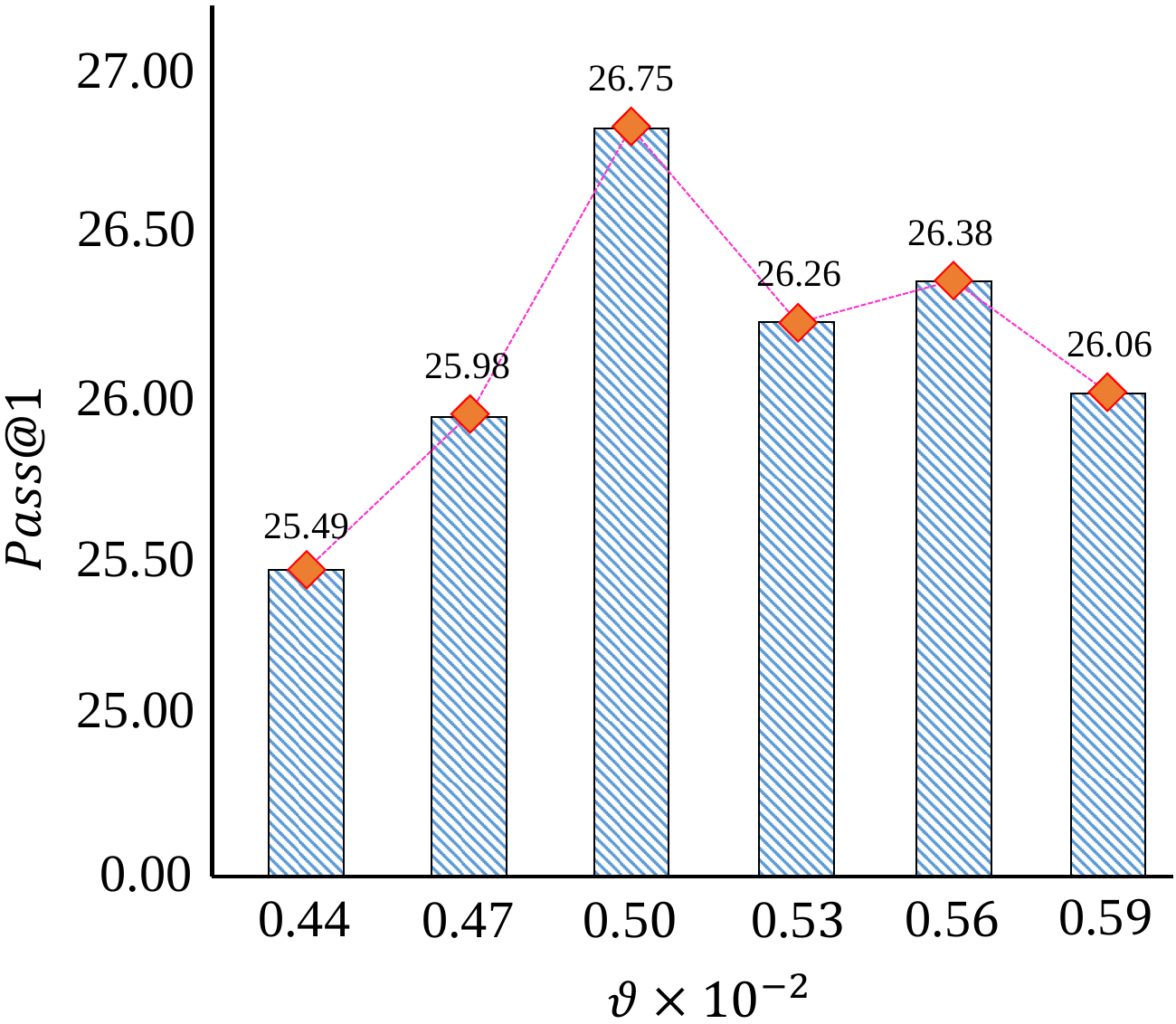}
    \caption{\textbf{\textit{Pass@$1$} scores of CodeLlama-7b~\cite{touvron2023llama} under different values of $\vartheta$}.}
    \label{fig:standard_deviation}
\end{figure}

\begin{table*}[!t]
  \centering
  \resizebox{1.0\linewidth}{!}{
    \begin{tabular}{ccccccccccc}
    \toprule
    \multirow{2}[4]{*}{\textbf{Model}} & \multirow{2}[4]{*}{\textbf{Methods}} & \multicolumn{7}{c}{\textbf{\textit{Pass@$k$}}} & \multirow{2}[4]{*}{\textbf{Avg}} & \multirow{2}[4]{*}{$\boldsymbol{\Delta}\left(\uparrow\right)$} \\
\cmidrule{3-9}          &       & 1     & 3     & 5     & 8     & 10    & 12    & 15    &       &  \\
    \midrule
    \multicolumn{11}{c}{\textit{HumanEval}} \\
    \midrule
    \midrule
    \multirow{2}[1]{*}{Llama2-7b} & Standard & 10.23	& 17.02	& 20.30	& 23.44	& 25.03	& 26.40	& 28.14  & \textbf{21.51}  & -  \\
    & \qquad\textit{\textbf{\textasciitilde w/USCD}} & 10.56	& 17.79	& 21.39	& 25.11	& 27.08	& 28.81	& 31.10  & \textbf{23.12}  & \textcolor{red}{\textbf{7.48\%}}  \\ \hdashline
    \multirow{2}[0]{*}{CodeLlama-7b} & Standard & 25.97	& 41.68	& 49.59	& 57.00	& 60.42	& 63.04	& 65.85
    & \textbf{51.94}  & -  \\
    & \qquad\textit{\textbf{\textasciitilde w/USCD}}  & 26.75	& 42.28	& 49.96	& 57.52	& 61.35	& 64.52	& 68.29  & \textbf{52.95}  & \textcolor{red}{\textbf{1.94\%}}  \\ \hdashline
    \multirow{2}[0]{*}{StarCoder} & Standard & 28.40	& 45.10	& 52.93	& 59.80	& 62.93	& 65.38	& 68.30  & \textbf{54.69}  & -  \\
    & \qquad\textit{\textbf{\textasciitilde w/USCD}}  & 28.58	& 45.63	& 53.95	& 61.54	& 65.03	& 67.72	& 70.73  & \textbf{56.17}  & \textcolor{red}{\textbf{2.71\%}}  \\ \hdashline
    \multirow{2}[1]{*}{WizardCoder-15b} & Standard & 45.97	& 62.25	& 67.64	& 71.44	& 72.92	& 73.97	& 75.00  & \textbf{67.03}  & -  \\
    & \qquad\textit{\textbf{\textasciitilde w/USCD}}   & 47.20	& 64.61	& 71.00	& 75.87	& 77.88	& 79.37	& 81.10  & \textbf{71.00}  & \textcolor{red}{\textbf{5.92\%}} \\ \hdashline
    \multirow{2}[1]{*}{Incoder-6b}	& Standard & 9.07	& 15.37	& 18.95	& 22.73	& 24.73	& 26.46	& 28.66 & \textbf{20.85}  & -  \\
    & \qquad\textit{\textbf{\textasciitilde w/USCD}}   & 10.08	& 16.66	& 20.57	& 25.06	& 27.53	& 29.67	& 32.32  & \textbf{23.13}  & \textcolor{red}{\textbf{10.94\%}}  \\
    \midrule
    \multicolumn{11}{c}{\textit{MBPP}} \\
    \midrule
    \midrule
    \multirow{2}[1]{*}{Llama2-7b} & Standard & 
    11.16	& 22.16	& 28.16	& 33.90	& 36.72	& 39.07	& 42.00  & \textbf{30.45}  & -  \\
    & \qquad\textit{\textbf{\textasciitilde w/USCD}}   & 11.77	& 22.96	& 28.97	& 34.61	& 37.27	& 39.42	& 42.51  & \textbf{31.07}  & \textcolor{red}{\textbf{2.04\%}}  \\ \hdashline
    \multirow{2}[0]{*}{CodeLlama-7b} & Standard & 30.04	& 48.09	& 55.23	& 61.02	& 63.51	& 65.42	& 67.60 & \textbf{55.84}  & -  \\
    & \qquad\textit{\textbf{\textasciitilde w/USCD}}   & 30.48	& 48.18	& 55.21	& 61.09	& 63.75	& 65.87	& 68.40  & \textbf{56.14}  & \textcolor{red}{\textbf{0.54\%}}  \\ \hdashline
    \multirow{2}[0]{*}{StarCoder} & Standard & 34.69	& 52.45	& 59.13	& 64.44	& 66.69	& 68.41	& 70.04  & \textbf{59.41}  & -  \\
    & \qquad\textit{\textbf{\textasciitilde w/USCD}}   & 37.28	& 54.03	& 59.93	& 64.42	& 66.25	& 67.63	& 69.20  & \textbf{59.82}  & \textcolor{red}{\textbf{0.69\%}}  \\ \hdashline
    \multirow{2}[1]{*}{WizardCoder-15b} & Standard & 45.11	& 56.56	& 60.90	& 64.61	& 66.28	& 67.60	& 69.20  & \textbf{61.47}  & -  \\
    & \qquad\textit{\textbf{\textasciitilde w/USCD}}   & 45.76	& 57.51	& 62.14	& 65.93	& 67.58	& 68.87	& 70.40  & \textbf{62.20}  & \textcolor{red}{\textbf{1.19\%}}  \\ \hdashline
    \multirow{2}[1]{*}{Incoder-6b}	& Standard & 9.53	& 20.08	& 25.95	& 31.44	& 33.99	& 36.00	& 38.40  & \textbf{27.91}  & -   \\
    & \qquad\textit{\textbf{\textasciitilde w/USCD}}   & 12.05	& 24.09	& 30.60	& 36.57	& 39.29	& 41.40	& 43.80  & \textbf{32.54}  & \textcolor{red}{\textbf{16.59\%}}  \\
    \bottomrule
    \end{tabular}
    }
    \caption{\textbf{The performance of LLMs (e.g., Llama2-7b, CodeLlama-7b, etc) incorporating USCD mechanism on the HumanEval and MBPP benchmarks}. We set the MBPP benchmark following~\cite{fried2023incoder}, i.e., adding an input-output example after the text. The red colour shows the ratio of performance improvement achieved by using a USCD mechanism compared to using standard prompts. During the experiment, we use a temperature of $0.8$ and top-$p$=$0.95$.}
    \label{tab:HumanEval_and_MBPP}
\end{table*}%

\begin{table*}[!t]
  \centering
  \resizebox{1.0\linewidth}{!}{
    \begin{tabular}{ccccccccccc}
    \toprule
    \multirow{2}[4]{*}{\textbf{Model}} & \multirow{2}[4]{*}{\textbf{Methods}} & \multicolumn{7}{c}{\textbf{Multi-lingual HumanEval}}                   &       & \multirow{2}[4]{*}{\textbf{Avg}} \\
\cmidrule{3-10}          &       & C++   & JAVA  & PHP   & C\#    & Bash  & D     & Lua    & JavaScript    &  \\
    \midrule
    \multirow{2}[1]{*}{Llama2-7b} & Standard & 6.83  & 9.32  & 9.32  & 7.45  & 3.73  & 5.59  & 10.56  & 13.04  & \textbf{8.23}  \\
    & \qquad\textit{\textbf{\textasciitilde w/USCD}}   & 7.45  & 10.56  & 9.94  & 9.94  & 4.35  & 8.23  & 11.68  & 13.66  & \textbf{9.48}  \\ \hdashline
    \multirow{2}[0]{*}{CodeLlama-7b} & Standard & 26.71  & 31.68  & 21.12  & 21.12  & 12.42  & 11.18  & 16.77  & 31.68  & \textbf{21.59}  \\
    & \qquad\textit{\textbf{\textasciitilde w/USCD}}   & 29.81  & 32.91  & 22.45  & 24.22  & 11.80  & 15.38  & 25.47  & 34.16  & \textbf{24.53}  \\ \hdashline
    \multirow{2}[0]{*}{StarCoder} & Standard & 31.68  & 28.57  & 26.09  & 18.63  & 11.23  & 14.29  & 24.22  & 32.30  & \textbf{23.38}  \\
    & \qquad\textit{\textbf{\textasciitilde w/USCD}}   & 30.43  & 31.68  & 28.71  & 24.22  & 11.80  & 14.56  & 20.50  & 31.68  & \textbf{24.20}  \\ \hdashline
    \multirow{2}[1]{*}{WizardCoder-15b} & Standard & 38.51  & 35.40  & 39.75  & 27.95  & 14.91  & 12.42  & 26.71  & 39.75  & \textbf{29.43}  \\
    & \qquad\textit{\textbf{\textasciitilde w/USCD}}   & 42.24  & 36.65  & 39.75  & 28.57  & 18.01  & 13.03  & 28.57  & 43.48  & \textbf{31.29}  \\ \hdashline
    \multirow{2}[1]{*}{Incoder-6b} & Standard & 10.56  & 8.07  & 6.83  & 6.21  & 2.48  & 1.86  & 4.35  & 8.70  & \textbf{6.13}  \\
    & \qquad\textit{\textbf{\textasciitilde w/USCD}}   & 13.04  & 9.94  & 9.32  & 6.83  & 3.73  & 2.56  & 5.59  & 7.45  & \textbf{7.31}  \\
    \bottomrule
    \end{tabular}
    }
   \caption{\textbf{The performance of LLMs based on USCD mechanism on the Multi-Lingual HumanEval benchmark}. During the experiment, we achieved a \textit{Pass@$1$} score using a temperature of $0.8$ and top-$p$=$0.95$.}
   \label{tab:Multi-lingual_HumanEval}%
\end{table*}%

\begin{figure}[!t]
    \centering
    \includegraphics[width=0.47\textwidth]{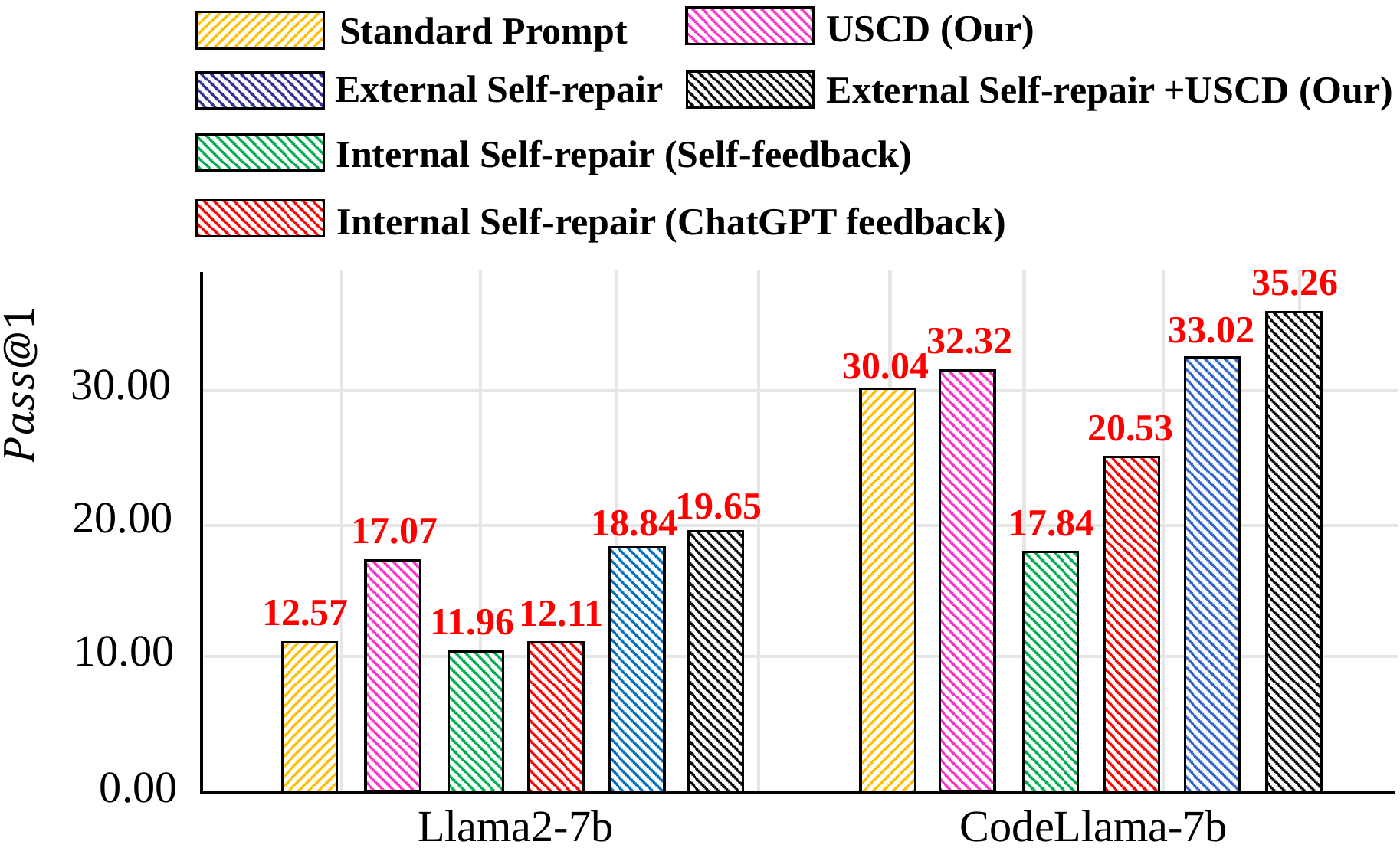}
    \caption{\textbf{Performance comparison of the used Llama2-7b~\cite{touvron2023llama} and CodeLlama-7b~\cite{rozière2023code}, using standard prompt, USCD mechanism, Internal Self-repair (Self-feedback), Internal Self-repair (ChatGPT~\cite{openai2023gpt4} feedback), External Self-repair and External Self-repair + USCD mechanism on the HumanEval benchmark~\cite{chen2021evaluating}}. Internal Self-repair (Self-feedback) means evaluating the generated code by oneself to obtain feedback. Internal Self-repair (ChatGPT feedback) means using ChatGPT to evaluate the generated code and obtain feedback. External Self-repair refers to using an evaluator to identify and obtain feedback on erroneous code. We set up Internal Self-repair and External Self-repair following the methods outlined in~\cite{huang2023large, valmeekam2023can} and~\cite{olausson2023self}, respectively. During the experiment, we use a temperature of $0.1$ and top-$p$=$0.95$.}
    \label{fig:uscd_vs_repair}
\end{figure}

\subsection{Ablation Studies}
\textbf{The impact of input-output examples $\boldsymbol{d}$ on code generation by LLMs}. In Section~\ref{sec:noise_prompt_construction} we simply demonstrated the performance of LLMs in code generation without input-output examples $\boldsymbol{d}$ from a quantitative perspective. Next, we thoroughly analyze the impact of input-output examples $\boldsymbol{d}$ on code generation by LLMs. We select the CodeLlama-7b model for testing using the HumanEval benchmark, and the results are shown in Table~\ref{tab:different_input-output_example}. We can find that: 1) When An input-output example randomly is removed from Standard Prompt, the code generation performance of CodeLlama-7b dramatically decreases; 2) As the input-output examples in the standard prompt gradually decrease, the score of CodeLlama-7b also gradually decreases. The above findings once again demonstrate that without input-output examples, LLMs can generate more noise (i.e., incorrectly code tokens) during the code generation process, resulting in lower scores.

% in-depth
\noindent
\textbf{The impact of input-output examples $\boldsymbol{d}$ on USCD mechanism}. We have analyzed how input-output examples $\boldsymbol{d}$ affect the generation of code by LLMs. Now, we will gradually reduce the standard prompt of input-output examples $\boldsymbol{d}$ and use it as the lame prompt for USCD mechanism experiments, as shown in Table~\ref{tab:different_input-output_example_contrastive_decoding}. The results indicate that the fewer input-output examples $\boldsymbol{d}$ in the lame prompt, the better the performance of the LLMs. This also shows that a lame prompt without input-output examples can serve as an effective negative prompt in contrastive decoding.

\noindent
\textbf{Role of coefficient $\rho$}. 
We keep other parameters consistent, i.g., $\vartheta=0$, and analyze the impact of $\rho$, as shown in Table~\ref{tab:coefficient_p}. We can observe that: 1) when the coefficient $\rho$ is smaller, the scores of Incoder-7b, CodeLlama-7b, and StarCoder models are essentially consistent with directly using the standard prompts. This also indicates that the role of the USCD mechanism is limited at this point. 2) as the coefficient $\rho$ increases, the scores of Incoder-7b, CodeLlama-7b, and StarCoder models decreases. It indicates that the USCD mechanism not only fails to improve but also introduces more noise. Therefore, we need $\rho$ to be within a certain range to unleash the maximum potential of the USCD mechanism.

\noindent
\textbf{Role standard deviation $\vartheta$}. In this part, we conducted experiments related to standard deviation, analyzing its impact as shown in Figure~\ref{fig:standard_deviation}. We can observe that: 1)
When the value of $\vartheta$ is set too large, it may cause the normal output distribution of the standard prompts to be incorrectly repaired by the USCD mechanism, leading to a gradual decrease in the score of the generated code by CodeLlama-7b; 2) When $\vartheta$ is set too small, the output distribution with noise in the standard prompts may not be repaired, resulting in a lower score for CodeLlama-7b as well. Therefore, we need to carefully adjust the value of $\vartheta$ to ensure it falls within an appropriate range so that the USCD mechanism can work.

\subsection{Main Results}
We validated the one-pass code generation quality of the improved LLMs using the USCD mechanism under $\vartheta=0.5\times10^{-2}$ and $\rho=0.3$. The experimental results are shown in Figure~\ref{fig:uscd_vs_repair}, Table~\ref{tab:HumanEval_and_MBPP} and Table~\ref{tab:Multi-lingual_HumanEval}. We have derived the following three conclusions.

\noindent
\textbf{Compared to self-repair methods, the USCD mechanism is highly competitive}. In Figure~\ref{fig:uscd_vs_repair}, we can observe: 1) The External Self-repair of LLMs improves the quality of code generation. This also indicates that Llama2-7b and CodeLlama-7b are capable of fixing erroneous code.
2) The Internal Self-repair of LLMs do not achieve the desired improvement. This indicates that Llama2-7b and CodeLlama-7b are unable to obtain feedback on errors, thereby failing to achieve successful repairs.
3) The USCD mechanism can be effectively combined with Self-repair methods to enhance the code generation quality of LLMs. For instance, the combination of the USCD mechanism with External Self-repair methods using Llama2-7b and CodeLlama-7b improved performance on the HumanEval benchmark by 4.30\% and 6.78\%, respectively, compared to using only external Self-repair methods.

\noindent
\textbf{The USCD mechanism in multiple programming languages can significantly improve the generated code results}. In Table~\ref{tab:HumanEval_and_MBPP}, and~\ref{tab:Multi-lingual_HumanEval}, compared to directly using standard prompts, both code-specialized and general models have shown significant improvements with the introduction of a USCD mechanism in multiple programming languages. Specifically, Llama2-7b has improved by 7.48\%, and 15.19\% on the HumanEval and Multi-lingual benchmarks, respectively. Incoder-6b has seen improvements of 10.94\%, and 19.25\% on the HumanEval and Multi-lingual benchmarks, respectively. CodeLlama-7b, StarCoder, and WizardCoder-15b also show significant improvements. It can be shown that the use of the USCD mechanism can improve some wrongly predicted tokens in the process of code generation, so that high-quality code can be generated. In addition, during the generation process, the USCD mechanism does not require external feedback or the use of an evaluator.

\noindent
\textbf{With a standard prompt consisting of an input-output example, the USCD mechanism can also make significant improvements}. In the MBPP benchmark, LLMs often struggle to generate good code with only an input-output example prompt. However, integrating the USCD mechanism into LLMs yields significant improvements compared to standard prompts. In the MBPP benchmark, Llama2-7b and Incoder-6b achieved improvements of 2.04\% and 16.59\%, respectively. Other LLMs also exhibit noticeable improvements, e.g., CodeLama-7b, StarCoder, etc. Results show our USCD also significantly improved the standard prompt of input-output examples. 

\section{Related Work}
\label{sec:related_work}
\subsection{Code Generation of LLMs} 
Existing code generation methods can be mainly divided into four types: code generation methods based on code features~\cite{ling2016latent, yin2017syntactic, rabinovich2017abstract}, combined external search code generation methods~\cite{hayati2018retrieval, hashimoto2018retrieve, guo2019coupling}, post-processing based code generation methods~\cite{jain2022jigsaw, wang2022compilable, le2022coderl}, and in-context prompting methods~\cite{li2023structured,ahmed2024automatic,li2024acecoder}.
% The code generation task refers to the process where a machine generates specific programming language code snippets based on the developer's requirements, which are written in natural language and include input and output samples, environment information, etc. In some methods, a post-processing step is added to ensure the generated code's executability. The code completion task involves automatically understanding the developer's intentions based on the context of the existing code during the coding process, and completing the code for them using a code completion algorithm model.
% Existing code generation methods can be mainly divided into three categories: code generation methods based on code features~\cite{ling2016latent, yin2017syntactic, rabinovich2017abstract}, combined external search code generation methods~\cite{hayati2018retrieval, hashimoto2018retrieve, guo2019coupling}, and post-processing based code generation methods~\cite{jain2022jigsaw, wang2022compilable, le2022coderl}.

The code generation method based on code features~\cite{ling2016latent, yin2017syntactic, rabinovich2017abstract} is to learn natural language features from the training data and realize the conversion between natural language and code features, e.g., Ling et al~\cite{ling2016latent} used natural language descriptions of the abilities or effects of a card to automatically generate the corresponding card definition code (i.e., Java and Python) to reduce the time cost of card effect development.

The approach to generating code through external retrieval~\cite{hashimoto2018retrieve, guo2019coupling} involves aiding the decoder in code generation by fetching similar code, thereby diminishing the decoding space and ultimately improving the quality of the generated code. As the model can access external knowledge through retrieval to supplement the gaps in its information, the combination of code generation with retrieval is more aligned with the practices of the majority of developers.

Post-processing methods~\cite{jain2022jigsaw, wang2022compilable} in code generation often involve testing the model using test cases, and offering feedback on the generation process and outcomes to enhance the quality of the code. Some researchers~\cite{le2022coderl} also directly employ test cases to fortify the model during its training phase, which in turn, enhances the quality of the generated code.

The in-context prompting methods~\cite{li2023structured,ahmed2024automatic,li2024acecoder} usually involves adding relevant instructions and examples to the original standard prompt, guiding the LLM to generate a series of reasoning steps that generate the final code., e.g., Li et al~\cite{li2024acecoder} enhanced the code generation performance of an LLM by retrieving examples from the training set that align with the current standard prompt.

% Appendix~\ref{sec:code_generation} provides a detailed description of code generation methods based on code features, combined external search code generation methods, and post-processing based code generation methods. 
Unlike the above four types of methods, the proposed USCD mechanism neither requires pre-training or fine-tuning models nor retrieving external knowledge and post-processing operations. Instead, the USCD mechanism utilizes standard prompts to construct lame prompts for USCD operations to eliminate the noise existing in one-pass code generation.

\subsection{Contrastive Decoding}
Contrastive decoding~\cite{li2023contrastive} is an effective test-time strategy to reduce predictive errors by 1) designing positive and negative prompt and 2) subtracting the output distribution of the negative prompt from the output distribution of the positive prompt.
Existing work directly employs contrastive decoding to enhance text generation quality~\cite{chia2023contrastive,shi2023trusting}, safety~\cite{zhong-etal-2024-rose}, and reducing translation errors~\cite{sennrich2023mitigating}. In addition, some studies have applied contrastive decoding to multimodal visual recognition to alleviate visual hallucinations~\cite{leng2023mitigating,wang-etal-2024-mitigating}.

Unlike existing methods, we mainly perform selective contrastive decoding on uncertain noise in the standard prompt to improve the quality of one-pass generated code.

\begin{table*}[!t]
  \centering
  \resizebox{1.0\linewidth}{!}{
    \begin{tabular}{cccccccccccc}
    \toprule
    \multicolumn{1}{c}{Model} & \multicolumn{10}{c}{CodeLlama-7b} \\
    \midrule
    \multirow{2}[1]{*}{Entropy} & $\vartheta \times 10^{-1}$     & $0.26$  & $0.29$  & $0.32$  & $0.35$  & $0.38$  & $0.41$  & $0.44$  & $0.47$ & $0.50$ & $0.53$\\
          & \textit{Pass@$1$} & $25.04$  & $26.02$  & $24.55$  & $25.61$  & $25.89$  & $\textcolor{red}{\textbf{26.02}}$  & $25.37$  & $24.85$ & $24.23$ & $23.89$ \\ \hdashline
    \multirow{2}[0]{*}{Quartiles} & $\vartheta \times 10^{-9}$     & $0.34$  & $0.36$  & $0.38$  & $0.40$  & $0.42$  & $0.44$  & $0.46$  & $0.48$ & $0.50$ & $0.52$\\
          & \textit{Pass@$1$} & $23.74$  & $22.64$  & $22.52$  & $23.29$  & $\textcolor{red}{\textbf{24.27}}$  & $22.64$  & $22.24$  & $21.95$ & $21.62$ & $21.31$ \\ \hdashline
    \multirow{2}[1]{*}{Standard deviation} & $\vartheta \times 10^{-2}$     & $0.38$  & $0.41$  & $0.44$  & $0.47$  & $0.50$  & $0.53$  & $0.56$  & $0.59$ & $0.62$ & $0.65$ \\
          & \textit{Pass@$1$} & $25.06$  & $25.23$  & $25.49$  & $25.98$  & $\textcolor{red}{\textbf{26.75}}$  & $26.26$  & $26.38$  & $25.91$ & $25.53$ & $25.14$ \\
    \bottomrule
    \end{tabular}%
   }
  \caption{\textbf{The performance of CodeLlama-7b using entropy, quartiles, and standard deviation for pre-judgment on HumanEval benchmark}. The red colour shows the best result.}
  \label{tab:entropy_quartiles_standard}%
\end{table*}%

\begin{figure}[!t]
    \centering
    \includegraphics[width=0.47\textwidth]{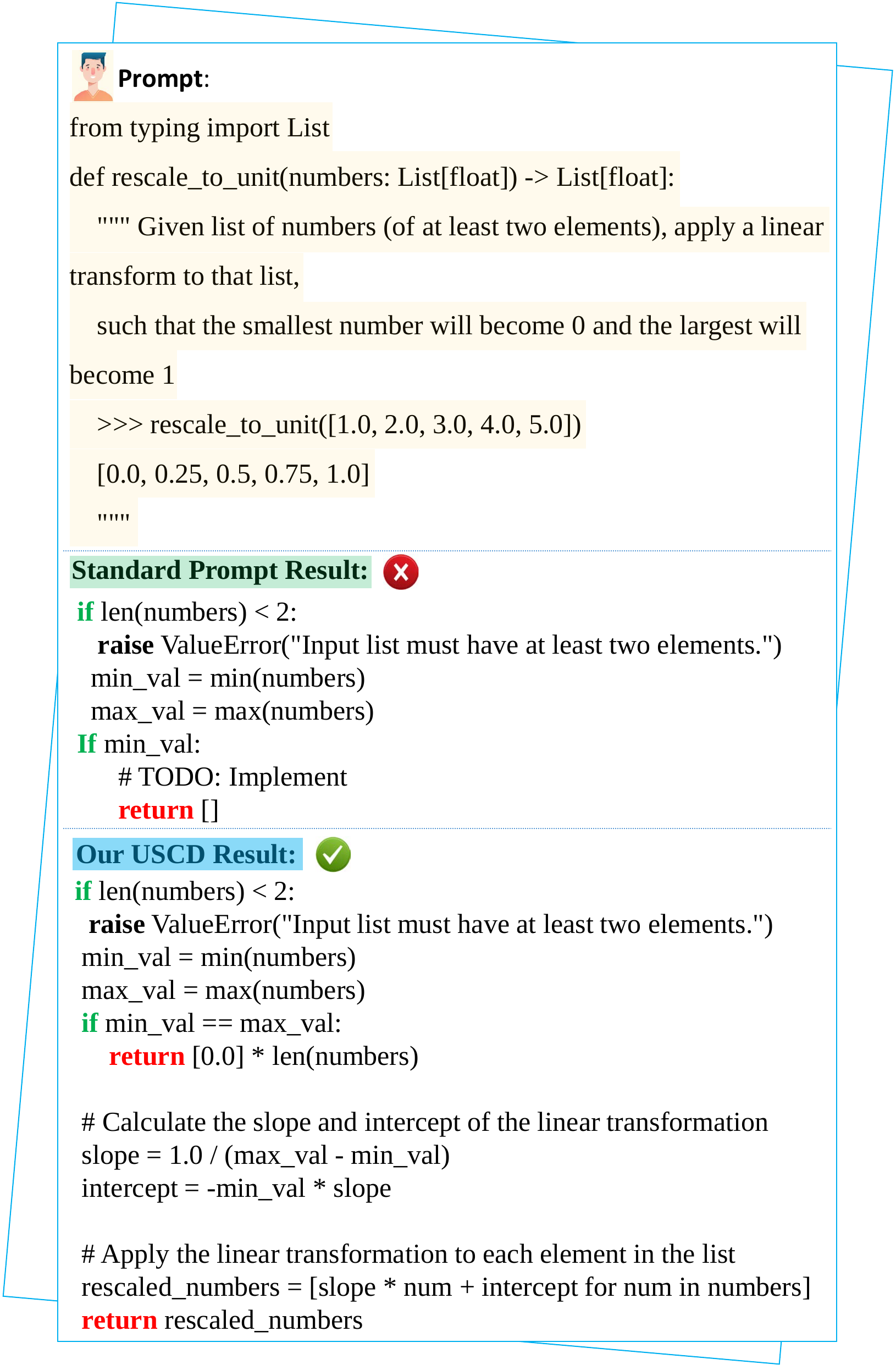}
    \caption{\textbf{CodeLlama-7b~\cite{touvron2023llama} directly uses standard prompt and USCD mechanism for case study on HumanEval benchmark~\cite{chen2021evaluating}}. The results generated by the standard prompt failed during testing, but the results generated by our USCD mechanism passed the tests successfully.}
    \label{fig:case_study}
\end{figure}

\section{Discussion}

Here we discuss why we use the standard deviation as the prediction criterion and show the detailed effects of USCD through several case studies.

\noindent
\textbf{Why choose standard deviation as a pre-judgment criterion}?
We improve the output distribution of standard prompts by using the USCD mechanism. This distribution is a discrete distribution of a set of data. Therefore, metrics for measuring the degree of continuous distribution changes and describing the state of discrete distribution, e.g., entropy and quartiles, are not appropriate. We use standard deviation, entropy, and quartiles as pre-judgments while keeping other parameters consistent, for the corresponding experiments, as shown in Table~\ref{tab:entropy_quartiles_standard}. From the experimental performance of standard deviation, entropy, and quartiles shown in Table~\ref{tab:entropy_quartiles_standard}, we observe that using standard deviation to measure the degree of variation in the current output distribution is more appropriate.

\noindent
\textbf{Case studies}. To better observe the improvement in code quality generated using the USCD mechanism compared to directly using the standard prompt, we show the results of code generation using the USCD mechanism and the standard prompt in Figure~\ref{fig:case_study} is located in the Appendix). We can find that directly using the standard prompt in the one-pass code generation process will incorrectly predict ``If'', leading to a lower quality of the generated code subsequently. However, our USCD mechanism can eliminate the prediction deviation in the generation process of standard prompts by using lame prompts, ensuring the subsequent generation of good code quality. 

\section{Conclusion}
\label{sec:conclusion}
To improve the one-pass code generation performance for LLMs, and reduce the impact of output noise, we propose a novel uncertainty-aware selective contrastive decoding (USCD) mechanism. This mechanism first pre-judges whether there is noise in the output distribution of standard prompts using the standard deviation. Then, it uses a lame prompt to eliminate noise in the output distribution of standard prompts and enhance the quality of code generation. Moreover, this mechanism is highly flexible and versatile. We further discuss why we chose standard deviation as the prediction and use a case study to visually demonstrate the improvement effects of the USCD mechanism.

\section*{Limitations}
\label{sec:limitations}
Although our USCD can improve the results of one-pass code generation, there are also some limitations to this mechanism: 1) The process of using the USCD mechanism obstructs the decoding time; 2) For some proprietary LLMs (e.g., ChatGPT) that utilize API interfaces, the USCD mechanism is not applicable. In the future, we will propose more advanced decoding mechanisms to improve the quality of one-pass code generation by LLMs and to accelerate the inference speed of LLMs.

\section*{Ethics Statement} We take ethical considerations very seriously and strictly adhere to the ACL Ethics Policy. This paper proposes an USCD mechanism to improve one-pass code generation in the context of LLMs. All employed models and datasets in this paper are publicly available and have been widely adopted by researchers. All experimental results upon these open models and datasets are reported accurately and objectively. Thus, we believe that this research will not pose any ethical issues.

\bibliography{anthology,custom}
\bibliographystyle{acl_natbib}
\appendix
% \newpage
\clearpage
\begin{table*}[!t]
  \centering
  \resizebox{1.0\linewidth}{!}{
    \begin{tabular}{cccccccccccccc}
    \toprule
    \multicolumn{1}{c}{\multirow{2}[4]{*}{\textbf{Benchmark}}} & \multicolumn{1}{c}{\multirow{2}[4]{*}{\textbf{Year}}} & \multicolumn{1}{c}{\multirow{2}[4]{*}{\textbf{Programming language}}} & \multicolumn{1}{c}{\multirow{2}[4]{*}{\textbf{Organization}}} & \multicolumn{9}{c}{\textbf{Input-output examples $\boldsymbol{C}$}} \\
\cmidrule{5-14}          &       &       &       & 0     & 1     & 2     & 3     & 4     & 5     & 6     & 7     & 8 & Total \\
    \midrule
    HumanEval & $2021$ & Python & OpenAI & $4$     & $28$    & $52$    & $37$    & $19$    & $12$ & $10$ & $1$ & $1$ & $164$ \\
    MBPP & $2021$ & Python & Google & $0$     & $500$   & $0$     & $0$     & $0$  & $0$ & $0$ & $0$   & $0$  & $500$ \\
    MultiPL-E & $2022$ & Multi-language & Northeastern University, USA & $4$     & $28$    & $52$    & $37$    & $19$    & $12$ & $10$ & $1$ & $1$ & $164^*$ \\
    \bottomrule
    \end{tabular}%
    }
  \caption{\textbf{The detailed description of HumanEval, MBPP, and MultiPL-E benchmarks}. We set the MBPP benchmark according to~\cite{fried2023incoder}, i.e., adding an input-output example after the text. ``$^*$'' shows the number of samples for each programming language.}
  \label{tab:humaneval_MBPP_MultiPL_example_number}
\end{table*}%

\begin{table*}[!t]
  \centering
  \resizebox{\linewidth}{!}{
    \begin{tabular}{cccccc}
    \toprule
    % \toprule
    \textbf{Model name} & \textbf{Organization}  & \multicolumn{1}{c}{\textbf{Years}} & \multicolumn{1}{c}{\textbf{Open-source}} & \multicolumn{1}{c}{\textbf{Task type}} &
    \multicolumn{1}{c}{\textbf{Source}} \\
    \midrule
    Llama2-7b & Meta & $2023$  & \CheckmarkBold & General & \url{https://huggingface.co/meta-llama/Llama-2-7b} \\
    CodeLlama-7b & Meta & $2023$  & \CheckmarkBold & Code-specialized & \url{https://huggingface.co/codellama/CodeLlama-7b-hf} \\
    StarCoder & Hugging Face   & $2023$  & \CheckmarkBold & Code-specialized & \url{https://huggingface.co/bigcode/starcoder} \\
    WizardCoder-15b & Microsoft  & $2023$  & \CheckmarkBold 
 & Code-specialized & \url{https://huggingface.co/WizardLMTeam/WizardCoder-15B-V1.0} \\
     Incoder-6b & Meta   & $2023$  & \CheckmarkBold 
 & Code-specialized & \url{https://huggingface.co/facebook/incoder-6B} \\
    \bottomrule
    % \bottomrule
    \end{tabular}%
    }
\caption{\textbf{Overview of the Evaluated Models}.}
  \label{tab:models_overview}%
\end{table*}%

\section{The Process of Constructing Lame Prompt}
\label{sec:lame_prompt}
\begin{figure}[!t]
    \centering
    \includegraphics[width=0.47\textwidth]{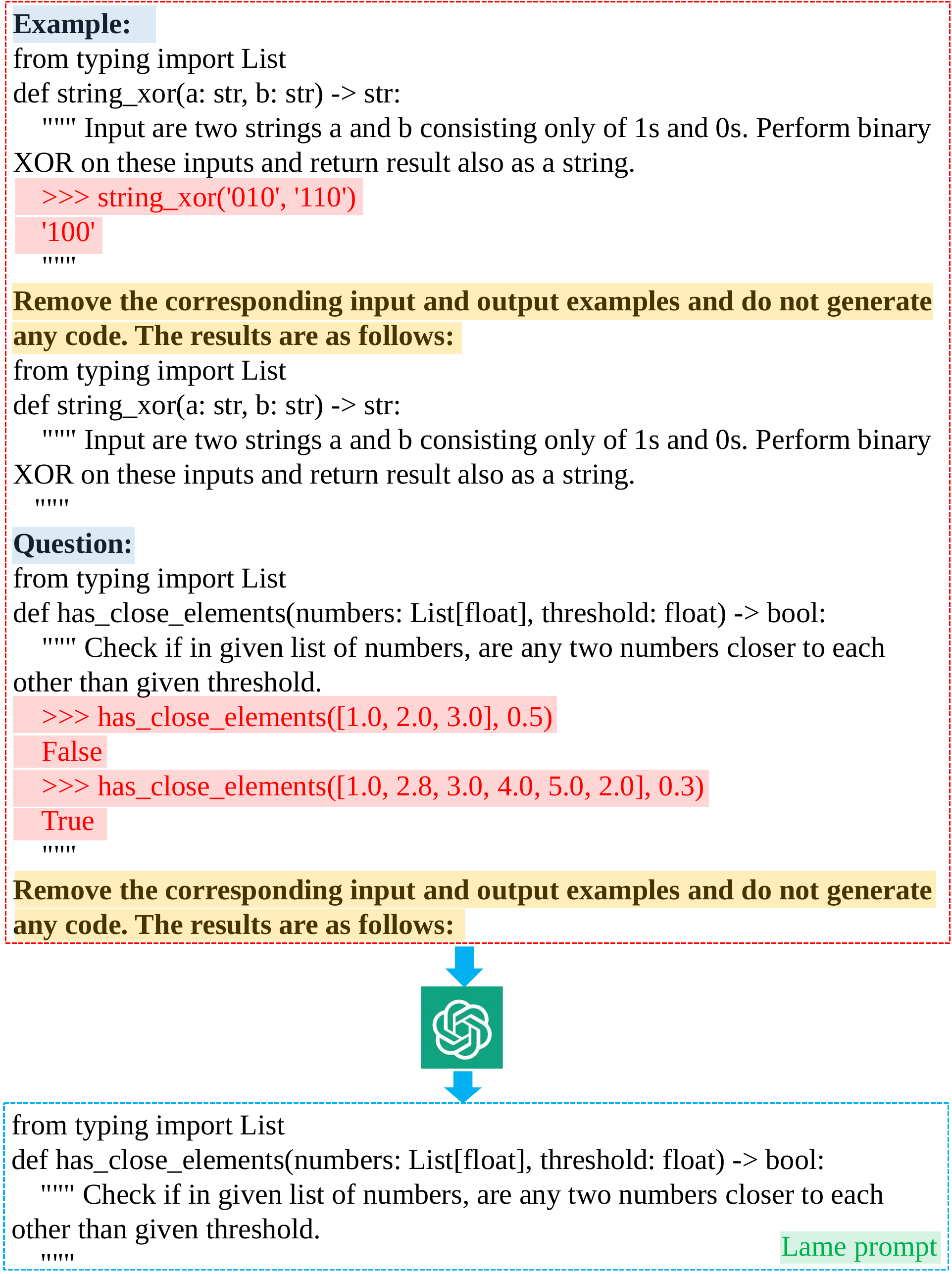}
    \caption{\textbf{The construction process of the lame prompt}.}
    \label{fig:lame_prompt_construction}
\end{figure}

According to the analysis in section II-2, we use a few-shot approach to have LLM (e.g., ChatGPT) remove the corresponding input-output examples, as illustrated in Figure~\ref{fig:lame_prompt_construction}.

\section{The Description of Test Benchmarks}
\label{sec:test_benchmark}
The HumanEval benchmark consists of $164$ handwritten Python programming problems and primarily focuses on language comprehension, algorithms, and basic mathematics. Additionally, the HumanEval benchmark mainly evaluates the function completion capability of LLMs.  Unlike the HumanEval benchmark, the MBPP benchmark primarily evaluates the function generation capability of LLMs. The test set for the MBPP benchmark consists of $500$ samples of Python language programs. MultiPL-E translates the HumanEval benchmark into eighteen other programming languages, e.g., C++, C\#, JAVA, PHP, and Bash. In this work, we selected eight commonly used programming languages (C++, JAVA, PHP, C\#, Bash, D, Lua, and JavaScript) based on the rankings from the TIOBE~\footnote{\url{https://www.tiobe.com/tiobe-index/}} leaderboard.

\section{The details of LLMs}
\label{sec:model}
We select general models, e.g., Llama2-7b~\cite{touvron2023llama} and code-specialized models, e.g., CodeLlama-7b~\cite{rozière2023code}, StarCode~\cite{li2023starcoder}, WizardCoder-15b~\cite{luo2023wizardcoder}, Incoder-6b~\cite{fried2023incoder}).

\noindent
\textbf{Llama2-7b~\cite{touvron2023llama}}. The Llama2-7b model, released by the Meta research team in July $2023$, is pre-trained with a parameter architecture of $70$ billion.

\noindent
\textbf{CodeLlama-7b~\cite{rozière2023code}}. The CodeLlama-7b model is fine-tuned based on the Llama model, primarily designed for tasks, e.g., code generation and code understanding. 

\noindent
\textbf{StarCoder~\cite{li2023starcoder}}. The StarCoder model is a $15.5$ billion parameter model trained using over $80$ programming languages from Stack (v1.2)~\footnote{\url{https://huggingface.co/datasets/bigcode/the-stack}}. 

\noindent
\textbf{WizardCoder~\cite{luo2023wizardcoder}}. WizardCoder is fine-tuned by applying the Evol-Instruct~\cite{xu2023wizardlm} method to Code LLMs.

\noindent
\textbf{Incoder-6b~\cite{fried2023incoder}}. Incoder-6b is trained on code using a causal-masked objective.

% \section{Code Generation of LLMs}
% \label{sec:code_generation}

\end{document}